\documentclass[seceq]{ptptex}

\notypesetlogo                       

\title{
Vector Manifestation in Hot and/or Dense Matter~\footnote{%
Talk given at YITP-RCNP Workshop
``Chiral Restoration in Nuclear Medium'' (October 7-9,2002).
\par
  This talk is based on the works done in 
  Refs.~\citen{HY:VM,HS:VMT,HKR,HKRS,HS:review}.
}
}

\author{
Masayasu \textsc{Harada}%
}

\inst{
Department of Physics, Nagoya University, Nagoya 464-8602, JAPAN\\
School of Physics, Seoul National University, Seoul 151-742, KOREA
}

\abst{
In this talk I summarize main features of the vector
manifestation (VM) which was recently proposed
as a novel manifestation of the Wigner realization of 
chiral symmetry in which the symmetry is restored at the critical
point by the massless degenerate pion (and its flavor partners) and
the $\rho$ meson (and its flavor partners) as the chiral partner.
I show how the VM is realized in hot 
and dense QCD using the effective field theory of QCD based on the
hidden local symmetry.
}

\begin{document}

\maketitle

\section{Introduction}

It is a very interesting subject to study the 
phase structure of QCD and the change of hadron
properties in hot and/or dense matter.
[See, for reviews, e.g., Refs.~\citen{HatsudaKunihiro,Pisarski:95,%
Brown-Rho:96,HatsudaShiomiKuwabara,%
Rapp-Wambach:00,Wilczek,Brown-Rho:01b}.]
Especially, the light vector meson is important for analyzing the
dilepton spectra in experimental facilities such as
the BNL Relativistic Heavy Ion Colider (RHIC).
In Refs.~\citen{Brown-Rho:91} and \citen{Brown-Rho:96} 
it was proposed that the
$\rho$-meson mass scales like the pion decay constant and vanishes at
the chiral phase transition point in hot and/or dense matter
(Brown-Rho scaling).

Recently, in Ref.~\citen{HY:VM}, 
the vector manifestation (VM) was proposed
as a novel manifestation of the Wigner realization of 
chiral symmetry in which the symmetry is restored at the critical
point by the massless degenerate $\pi$ 
(pion and its flavor partners) and $\rho$
($\rho$ meson and its flavor partners) as the chiral partner, in
sharp contrast to the traditional manifestation \'a la the linear
sigma model where the symmetry is restored by the degenerate pion and
the scalar meson.
It was shown that the VM is realized in the large flavor QCD by
using the hidden local symmetry (HLS)~\cite{BKUYY,BKY:88}
model which is an effective field
theory for $\pi$ and $\rho$ based on the chiral symmetry of QCD.
Then,
it was further shown that the VM can occur in the chiral restoration
in hot QCD~\cite{HS:VMT} and in dense QCD~\cite{HKR},
where an essential role was played by
the intrinsic temperature and density dependences of the
parameters of the HLS Lagrangian determined by extending the
Wilsonian matching between the HLS and the underlying 
QCD~\cite{HY:WM} to the one in hot and dense matter.
Moreover, several predictions from the VM in hot matter are made
in Refs.~\citen{HKRS,HS:review}.

In this talk I summarize main features of the VM,
especially compared with the conventional manifestation \'a la
the linear sigma model,
and then
show how the VM is realized in hot and/or dense matter
by formulating the VM in the effective field theory
for $\pi$ and $\rho$ based on the HLS.

This report is organized as follows:
In section~\ref{sec:VM}, I summarize a difference between the
VM and the conventional linear sigma model like manifestation
in terms of the chiral representations of low-lying mesons.
In section~\ref{sec:EFT},
I show the effective field theory of QCD based on the HLS
and present the renormalization group equations for the 
parameters of the HLS Lagrangian.
In section~\ref{sec:ITE},
I show the Wilsonian matching conditions which determine
the intrinsic temperature and/or density dependences of the bare 
parameters of the HLS Lagrangian
in terms of the parameters of the operator
product expansion in QCD, and derive the constraints on the  
parameters at the critical point.
Then, I show how the VM is realized in hot and dense matter
in section~\ref{sec:VMHDM}.
Finally, I give a brief summary in section~\ref{sec:summary}.
Several functions used in section~\ref{sec:VMHDM} are summarized
in Appendix~\ref{app:Functions}.

\section{Vector Manifestation}
\label{sec:VM}

In this section I briefly explain some features of the
vector manifestation (VM).
The VM was first proposed in
Ref.~\citen{HY:VM} as a novel manifestation of Wigner realization of
chiral symmetry where the vector meson $\rho$ becomes massless at the
chiral phase transition point. 
Accordingly, the (longitudinal) $\rho$ becomes the chiral partner of
the Nambu-Goldstone boson $\pi$.

The VM is characterized by
\begin{equation}
F_\pi^2 \rightarrow 0 \ , \quad
m_\rho^2 \rightarrow m_\pi^2 = 0 \ , \quad
F_\rho^2 / F_\pi^2 \rightarrow 1 \ ,
\end{equation}
where $F_\rho$ is the decay constant of 
(longitudinal) $\rho$ at $\rho$ on-shell.
This is completely different from the conventional picture based
on the linear sigma model (I call this GL manifestation after the
effective theory of Ginzburg--Landau or Gell-Mann--Levy.)
where the scalar meson becomes massless
degenerate with $\pi$ as the chiral partner.
Here, I
discuss the difference between the VM and the
GL manifestation in terms of the chiral representation of the mesons
by extending the analyses done in
Refs.~\citen{Weinberg:69,Gilman-Harari}.
Following Ref.~\citen{Weinberg:69},
I define the
axialvector coupling matrix $X_a(\lambda)$
by giving the matrix elements at zero invariant momentum transfer of
the axialvector current between states with collinear momenta as
\begin{equation}
\langle \vec{q} \,\lambda^\prime \, \beta \vert
( J_{5a}^{0} + J_{5a}^{3} )
\vert \vec{p} \,\lambda \, \alpha \rangle
=
\, 2 {\cal E} \, \delta_{\lambda\lambda^\prime}
\left[ X_a(\lambda) \right]_{\beta\alpha}
\ ,
\label{axial coupling}
\end{equation}
where $\alpha$ and $\beta$ are one-particle states
with momentum $\vec{p}$ and $\vec{q}$ in $3$ direction, 
$\lambda$ and $\lambda^\prime$ are their helicities.
The quantity ${\cal E}$ is defined by writing the condition of zero
invariant momentum transfer as
\begin{equation}
\vert \vec{p} \vert + \sqrt{ \vert\vec{p}\vert^2 + m_\alpha^2 }
=
\vert \vec{q} \vert + 
  \sqrt{ \vert\vec{q}\vert^2 + m_\beta^2 }
\equiv
{\cal E}
\ .
\end{equation}
It was stressed~\cite{Weinberg:69} that the definition
of the axialvector couplings in Eq.~(\ref{axial coupling}) can be 
used for particles of arbitrary spin, and in arbitrary collinear
reference frames, including both the frames in which $\alpha$ is at
rest and in which it moves with infinite momentum.
As was done in Ref.~\citen{Weinberg:69},
considering
the forward scattering process
$\pi_a + \alpha(\lambda) \rightarrow \pi_b + \beta(\lambda^\prime)$
and 
requiring the cancellation of the terms in the $t$-channel,
I obtain
\begin{equation}
\left[ \, X_a(\lambda)\,,\, X_b(\lambda) \,\right]
= i f_{abc} T_c
\ ,
\label{algebra}
\end{equation}
where
$T_c$ is the generator of $\mbox{SU}(N_f)_{\rm V}$ and 
$f_{abc}$ is the structure constant.
It should be noticed that Eq.~(\ref{algebra}) tells us that the
one-particle states of any given helicity must be assembled into
representations of chiral 
$\mbox{SU}(N_f)_{\rm L}\times\mbox{SU}(N_f)_{\rm R}$.
Furthermore, since Eq.~(\ref{algebra}) does not give any relations 
among the states with different helicities, those states can 
generally belong to the different representations even though
they form a single particle such as the longitudinal $\rho$
($\lambda=0$) and the transverse $\rho$ ($\lambda=\pm1$).
Thus, the notion of the chiral partners can be considered
separately for each helicity.

Let me consider the algebraic relation (\ref{algebra})
for zero helicity ($\lambda=0$) states
and saturate it by low-lying mesons;
the $\pi$, the (longitudinal) $\rho$, the (longitudinal) axialvector
meson denoted by $A_1$ ($a_1$ meson and its flavor partners)
and the scalar meson, and so on.
It should be noticed that, in the broken phase of chiral symmetry,
the Hamiltonian matrix defined by the matrix elements of the
Hamiltonian between states does not commute with the
axialvector coupling matrix.
Then, the algebraic representations of the 
axialvector coupling matrix do not coincide
with the mass eigenstates: There occur representation mixings.
Actually,
$\pi$ and the longitudinal $A_1$ 
are admixture of $(8\,,\,1) \oplus(1\,,\,8)$ and 
$(3\,,\,3^*)\oplus(3^*\,,\,3)$~\cite{Weinberg:69,Gilman-Harari}:
\begin{eqnarray}
\vert \pi\rangle &=&
\vert (3\,,\,3^*)\oplus (3^*\,,\,3) \rangle \sin\psi
+
\vert(8\,,\,1)\oplus (1\,,\,8)\rangle  \cos\psi
\ ,
\nonumber
\\
\vert A_1(\lambda=0)\rangle &=&
\vert (3\,,\,3^*)\oplus (3^*\,,\,3) \rangle \cos\psi 
- \vert(8\,,\,1)\oplus (1\,,\,8)\rangle  \sin\psi
\ ,
\label{mix pi A}
\end{eqnarray}
where the experimental value of the mixing angle $\psi$ is 
given by approximately 
$\psi=\pi/4$~\cite{Weinberg:69,Gilman-Harari}.  
On the other hand, the longitudinal $\rho$
belongs to pure $(8\,,\,1)\oplus (1\,,\,8)$
and the scalar meson to 
pure $(3\,,\,3^*)\oplus (3^*\,,\,3)$.

When the chiral symmetry is restored at the
phase transition point, the axialvector coupling matrix
commutes with the Hamiltonian matrix, and thus the 
chiral representations coincide with the mass eigenstates:
The representation mixing is dissolved.
{}From Eq.~(\ref{mix pi A}) one can easily see that
there are two ways to express the representations in the
Wigner phase of the chiral symmetry:
The conventional GL manifestation
corresponds to 
the limit $\psi \rightarrow \pi/2$ in which
$\pi$ is in the representation
of pure $(3\,,\,3^*)\oplus(3^*\,,\,3)$ 
together with the scalar meson,
while
the VM
to the limit $\psi\rightarrow 0$ in which the $A_1$ 
goes to a pure 
$(3\,,\,3^*)\oplus (3^*\,,\,3)$, now degenerate with
the scalar meson in the same representation, 
but not with $\rho$ in 
$(8\,,\,1)\oplus (1\,,\,8)$.
Namely, the
degenerate massless $\pi$ and (longitudinal) $\rho$ at the 
phase transition point are
the chiral partners in the
representation of $(8\,,\,1)\oplus (1\,,\,8)$.

\section{Effective Field Theory}
\label{sec:EFT}

In this section I show the effective field theory in which
the vector manifestation is formulated.
I should note that,
as is stressed in Ref.~\citen{HY:PRep},
the VM can be formulated only as a limit by approaching it from the
broken phase of chiral symmetry.
Then, for the formulation of the VM I need an effective field theory
(EFT)
including $\rho$ and $\pi$ 
in the broken phase which is not
necessarily applicable in the symmetric phase.
One of such EFTs is the model based on the 
hidden local symmetry (HLS)~\cite{BKUYY,BKY:88}
which includes $\rho$ as the gauge boson of the HLS in addition
to $\pi$ as the NG boson associated with the
chiral symmetry breaking in a manner fully consistent with
the chiral symmetry of QCD.
It should be noticed that,
in the HLS,
thanks to the gauge invariance
one can perform the
systematic chiral perturbation with including $\rho$
in addition to $\pi$.~\cite{Georgi,HY:PLB,Tanabashi,HY:WM,HY:PRep}
In subsection~\ref{ssec:HLS}, I will explain the model based
on the HLS, and then summarize the renormalization group equations for 
the parameters of the HLS Lagrangian in subsection~\ref{ssec:RGE}.

\subsection{Hidden Local Symmetry}
\label{ssec:HLS}

Let me describe the HLS model
based on the
$G_{\rm global} \times H_{\rm local}$ symmetry, where
$G = \mbox{SU($N_f$)}_{\rm L} \times 
\mbox{SU($N_f$)}_{\rm R}$  is the 
global chiral symmetry and 
$H = \mbox{SU($N_f$)}_{\rm V}$ is the HLS.
The basic quantities 
are the gauge boson 
$\rho_\mu$
and two 
variables 
\begin{eqnarray}
&&
\xi_{\rm L,R} = e^{i\sigma/F_\sigma} e^{\mp i\pi/F_\pi}
\ ,
\end{eqnarray}
where $\pi$
denotes the pseudoscalar NG boson
and $\sigma$~\footnote{
Note that this $\sigma$ is different from the scalar meson
in the linear sigma model.
}
the NG boson absorbed into $\rho_\mu$ 
(longitudinal $\rho$).
$F_\pi$ and $F_\sigma$ are relevant decay constants, and
the parameter $a$ is defined as
$a \equiv F_\sigma^2/F_\pi^2$.
The transformation properties of $\xi_{\rm L,R}$ are given by
\begin{eqnarray}
&&
\xi_{\rm L,R}(x) \rightarrow \xi_{\rm L,R}^{\prime}(x) =
h(x) \xi_{\rm L,R}(x) g^{\dag}_{\rm L,R}
\ ,
\end{eqnarray}
where $h(x) \in H_{\rm local}$ and 
$g_{\rm L,R} \in G_{\rm global}$.
The covariant derivatives of $\xi_{\rm L,R}$ are defined by
\begin{eqnarray}
&&
D_\mu \xi_{\rm L} =
\partial_\mu \xi_{\rm L} - i g \rho_\mu \xi_{\rm L}
+ i \xi_{\rm L} {\cal L}_\mu
\ ,
\nonumber\\
&&
D_\mu \xi_{\rm R} =
\partial_\mu \xi_{\rm R} - i g \rho_\mu \xi_{\rm R}
+ i \xi_{\rm R} {\cal R}_\mu
\ ,
\label{covder}
\end{eqnarray}
where $g$ is the HLS gauge coupling, and
${\cal L}_\mu$ and ${\cal R}_\mu$ denote the external gauge fields
gauging the $G_{\rm global}$ symmetry.

The HLS Lagrangian at the leading order
is given by~\cite{BKUYY,BKY:88}
\begin{equation}
{\cal L} = F_\pi^2 \, \mbox{tr} 
\left[ \hat{\alpha}_{\perp\mu} \hat{\alpha}_{\perp}^\mu \right]
+ F_\sigma^2 \, \mbox{tr}
\left[ 
  \hat{\alpha}_{\parallel\mu} \hat{\alpha}_{\parallel}^\mu
\right]
+ {\cal L}_{\rm kin}(\rho_\mu) \ ,
\label{Lagrangian}
\end{equation}
where ${\cal L}_{\rm kin}(\rho_\mu)$ denotes the kinetic term of
$\rho_\mu$ 
and
\begin{eqnarray}
&&
\hat{\alpha}_{\perp,\parallel}^\mu =
( D_\mu \xi_{\rm R} \cdot \xi_{\rm R}^\dag \mp 
  D_\mu \xi_{\rm L} \cdot \xi_{\rm L}^\dag
) / (2i)
\ .
\end{eqnarray}
When the kinetic term ${\cal L}_{\rm kin}(\rho_\mu)$
is ignored in the low-energy region,
the second term of Eq.~(\ref{Lagrangian}) vanishes
by integrating out $\rho_\mu$
and only the first term remains.
Then, 
the HLS model is reduced to the nonlinear sigma model based on $G/H$.

In Refs.~\citen{Brown-Rho:96,Brown-Rho:01b}, 
it was pointed out that the quasiquark picture is appropriate near
the phase transition point, and that quasiquark
mass approaches zero.
Then, 
following Refs.~\citen{HKR,HKRS},
I introduce the quasiquark field $\psi$ near the critical point
into the Lagrangian in addition to $\rho$ and $\pi$.
A chiral
Lagrangian for $\pi$ with the constituent quark (quasiquark) was
given in Ref.~\citen{MG}. 
In Ref.~\citen{BKY:88} the quasiquark
field $\psi$ is introduced into the HLS Lagrangian in such a
way that it transforms homogeneously under the HLS: 
$\psi\rightarrow h(x)\cdot \psi$ where $h(x) \in H_{\rm local}$. 
In Ref.~\citen{HKR}
the Lagrangian of Ref.~\citen{BKY:88} was extended 
to a general one
with which a systematic derivative expansion can be performed. 
Since
the model is introduced 
near the chiral phase transition
point where the quasiquark mass is expected to become small, 
the quasiquark mass $m_q$ is counted as
${\cal O}(p)$.
Furthermore, ${\cal O}(p)$ is assigned to the chemical
potential $\mu$ or the Fermi momentum $P_F$ as well as to the 
temperature $T$, as 
the cutoff is considered to be
larger than $\mu$ and $T$ even near the phase transition
point. 
By using this counting scheme 
the systematic
expansion can be performed
in the HLS with the quasiquark included. 
I should note
that this counting scheme is different from the one in the model
for $\pi$ and baryons given in Ref.~\citen{MOR} where the baryon
mass is counted as ${\cal O}(1)$. The leading order Lagrangian
including one quasiquark field and one anti-quasiquark field is
counted as ${\cal O}(p)$ and given by~\cite{HKR}
 \begin{eqnarray}
 \delta {\cal L}_{Q} &=& \bar \psi(x)( iD_\mu \gamma^\mu
       + \mu \gamma^0 -m_q )\psi(x)\nonumber\\
 &&+ \bar \psi(x) \left(
  \kappa\gamma^\mu \hat{\alpha}_{\parallel \mu}(x )
+ \lambda\gamma_5\gamma^\mu \hat{\alpha}_{\perp\mu}(x) \right)
        \psi(x) \label{lagbaryon}
 \end{eqnarray}
where $D_\mu\psi=(\partial_\mu -ig\rho_\mu)\psi$ and $\kappa$ and
$\lambda$ are constants to be specified later.

\subsection{Renormalization Group Equations}
\label{ssec:RGE}

At one-loop level
the Lagrangian 
(\ref{Lagrangian}) plus 
(\ref{lagbaryon}) generates the ${\cal O}(p^4)$
contributions including hadronic thermal/dense-loop effects as well as
divergent effects. The divergent contributions are renormalized by
the parameters, and thus the RGEs for three leading order
parameters $F_\pi$, $a$ and $g$ (and parameters of ${\cal O}(p^4)$
Lagrangian) are modified from those without quasiquark field. In
addition, I need to consider the renormalization group flow for
the quasiquark mass $m_q$~\footnote{The constants $\kappa$ and
  $\lambda$ will also run.
  The running will be small near the critical point, 
  so I will ignore their running here.
}. 
By 
calculating one-loop contributions for RGEs in an energy scale
${\cal M}$ for a given temperature $T$ and chemical potential $\mu$, 
the RGEs are
expressed as~\cite{HY:conformal,HKR,HS:review}
\begin{eqnarray}
{\cal M} \frac{dF_\pi^2}{d{\cal M}} &=& 
C\left[3a^2g^2F_\pi^2 +2(2-a){\cal M}^2 \right] 
-\frac{m_q^2}{2\pi^2}\lambda^2 N_c
\ ,
\nonumber\\
{\cal M} \frac{da}{d{\cal M}} &=&-C
(a-1) \left [3a(1+a)g^2-(3a-1)\frac{{\cal M}^2}{F_\pi^2} \right]
 +a\frac{\lambda^2}{2\pi^2}\frac{m_q^2}{F_\pi^2}N_c
\ ,
\nonumber\\
{\cal M}\frac{d g^2}{d {\cal M}}&=& -C\frac{87-a^2}{6}g^4
+\frac{N_c}{6\pi^2}g^4 (1-\kappa)^2
\ ,
\nonumber\\
{\cal M}\frac{dm_q}{d{\cal M}}
&=& -\frac{m_q}{8\pi^2}
\left[ (C_\pi-C_\sigma){\cal M}^2
  -m_q^2 (C_\pi-C_\sigma)
+M_\rho^2C_\sigma -4C_\rho \right]
\ ,
\label{rgewbaryons}
\end{eqnarray}
where $C = N_f/\left[2(4\pi)^2\right]$ and
\begin{equation}
C_\pi
\equiv
\frac{\lambda^2}{F_\pi^2} \frac{N_f^2 - 1}{2N_f}
\ ,
\quad
C_\sigma
\equiv
\frac{\kappa^2}{F_\sigma^2} \frac{N_f^2 - 1}{2N_f}
\ ,
\quad
C_\rho
\equiv
g^2 (1-\kappa)^2 \frac{N_f^2 - 1}{2N_f}
\ .
\end{equation}
It should be noted that the point $(g\,,\,a\,,\,m_q)=(0,1,0)$
is the fixed point of 
the RGEs in Eq.~(\ref{rgewbaryons})
which plays an essential role to realize the VM in the following
analysis of the chiral restoration in hot and/or dense QCD.

\section{Wilsonian Matching and Intrinsic Temperature/Density
Dependence} 
\label{sec:ITE}

The Wilsonian matching proposed in Ref.~\citen{HY:WM} is done by
matching 
the axialvector and vector current correlators derived from the
HLS with those by the 
operator product expansion (OPE) in
QCD at the matching scale $\Lambda$.
This was extended to non-zero temperature~\cite{HS:VMT}
and density~\cite{HKR}, and it was shown that the parameters of the
HLS Lagrangian have the intrinsic temperature and/or 
density dependences.
In this section I summarize the Wilsonian matching conditions
and the resultant conditions (VM conditions) for the bare 
parameters of the HLS
at the phase transition point.

In general
there is no longer Lorentz symmetry
in hot and/or dense matter,
and
the Lorentz non-scalar operators such as
$\bar{q}\gamma_\mu D_\nu q$ may exist in 
the form of the current correlators derived by the 
OPE. [see, e.g., Ref.~\citen{HKL}]
However, we neglect these contributions
since they give a small correction compared with the main term of
the form $1 + \frac{\alpha_s}{\pi}$.
In this approximation
the axialvector and vector correlators in the OPE are expressed
by the same form as used in Ref.~\citen{SVZ} with putting possible
temperature and/or density
dependences on the gluonic and quark
condensates:
\begin{eqnarray}
 G^{\rm{(QCD)}}_A (Q^2;T,\mu) 
&=&
 \frac{1}{8{\pi}^2}
 \Bigl[ 
   -     \bigl( 1 + \frac{\alpha _s}{\pi} \bigr)
                      \ln \frac{Q^2}{{\cal M}_Q^2} 
\nonumber\\ 
&& {}+ \frac{{\pi}^2}{3}
      \frac{\langle \frac{\alpha _s}{\pi}
       G_{\mu \nu}G^{\mu \nu} \rangle_{T,\mu} }{Q^4} +
      \frac{{\pi}^3}{3} \frac{1408}{27}
      \frac{\alpha _s{\langle \bar{q}q
      \rangle }^2_{T,\mu} }{Q^6}
 \Bigr] \ , 
\nonumber\\
    G^{\rm{(QCD)}}_V (Q^2;T,\mu) 
&=& \frac{1}{8{\pi}^2}
  \Bigl[ -
    \bigl( 1 + \frac{\alpha _s}{\pi} \bigr)
    \ln \frac{Q^2}{{\cal M}_Q^2} 
\nonumber\\
&& {}+\frac{{\pi}^2}{3}
      \frac{\langle \frac{\alpha _s}{\pi}
      G_{\mu \nu}G^{\mu \nu} \rangle_{T,\mu} }{Q^4} -
      \frac{{\pi}^3}{3} \frac{896}{27}
      \frac{\alpha _s{\langle \bar{q}q
      \rangle }^2_{T,\mu} }{Q^6}
 \Bigr] \ ,
\label{correlator OPE}
\end{eqnarray}
where ${\cal M}_Q$ is the renormalization point of QCD.
Consistently with the approximation adopted for the above current
correlators in the OPE,
I use the Lorentz invariant form of the HLS Lagrangian.
It should be noted
that the quasiquark does not contribute to the current
correlators at bare level.
Then,
the axialvector and the vector current
correlators in the HLS around the matching scale $\Lambda$ 
are well described by the same forms as those at 
$T=\mu=0$ with the bare 
parameters having the intrinsic temperature and/or density
dependences:
\begin{eqnarray}
   G^{\rm{(HLS)}}_A (Q^2;T,\mu) 
&=& \frac{F^2_\pi (\Lambda ;T,\mu)}{Q^2} -
             2z_2(\Lambda ;T,\mu)\ ,
\nonumber\\
   G^{\rm{(HLS)}}_V (Q^2;T,\mu) 
&=& \frac{F^2_\sigma (\Lambda ;T,\mu)[1 -
            2g^2(\Lambda ;T,\mu)z_3(\Lambda ;T,\mu)]}
                        {{M_\rho}^2(\Lambda ;T,\mu) + Q^2} -
                       2z_1(\Lambda ;T,\mu) \ .
\label{correlator HLS}
\end{eqnarray}
The Wilsonian matching conditions are obtained by setting
the above
correlators to be equal 
to those in Eq.~(\ref{correlator OPE})
at $\Lambda$
up until the first derivative.
Then the resultant forms of the 
Wilsonian matching conditions
at non-zero temperature and/or non-zero density
are expressed as~\footnote{%
  One might think that there appear corrections from $\rho$ and/or
  $\pi$ loops in the left-hand-sides of Eqs.~(\ref{eq:WMC A}) and
  (\ref{eq:WMC V}).
  However, such corrections are of higher order in the present
  counting scheme, and thus I neglect them here
  at $Q^2 \sim {\Lambda}^2$.
  In the low-energy scale
  I incorporate the loop effects into the correlators through the
  renormalization group equations.
}
\begin{eqnarray}
&&
   \frac{F^2_\pi (\Lambda ;T,\mu)}{{\Lambda}^2} - 
            \frac{F^2_\sigma (\Lambda ;T,\mu)[1 -
              2g^2(\Lambda ;T,\mu)z_3(\Lambda ;T,\mu)]}
             {{M_\rho}^2(\Lambda ;T,\mu) + {\Lambda}^2} 
\nonumber\\
&& \qquad\quad
  {} - 2\left[z_2(\Lambda ;T,\mu) - z_1(\Lambda ;T,\mu)\right] 
\nonumber\\
&& \qquad
     = \frac{32\pi}{9} \frac{\alpha _s{\langle \bar{q}q
      \rangle }^2_{T,\mu} }{{\Lambda}^6}
\ ,
\\
&&
   \frac{F^2_\pi (\Lambda ;T,\mu)}{{\Lambda}^2} = 
  \frac{1}{8{\pi}^2}
  \left[
    1 + \frac{\alpha _s}{\pi} 
    + \frac{2{\pi}^2}{3} 
    \frac{\langle \frac{\alpha _s}{\pi} 
      G_{\mu \nu}G^{\mu \nu} \rangle_{T,\mu} }{{\Lambda}^4} 
    + {\pi}^3 \frac{1408}{27}
      \frac{\alpha _s{\langle \bar{q}q \rangle }^2_{T,\mu} }
                                     {{\Lambda}^6}
  \right]
\ , \label{eq:WMC}
\label{eq:WMC A}
\\
&&
 \frac{F^2_\sigma (\Lambda ;T,\mu)}{{\Lambda}^2}
 \frac{{\Lambda}^4[1 - 2g^2(\Lambda ;T,\mu)z_3(\Lambda ;T,\mu)]}
     {({M_\rho}^2(\Lambda ;T,\mu) + {\Lambda}^2)^2}
\nonumber\\
&& \qquad
   = \frac{1}{8{\pi}^2}
   \left[ 1 + \frac{\alpha _s}{\pi}
       + \frac{2{\pi}^2}{3} \frac{\langle \frac{\alpha _s}{\pi}
         G_{\mu \nu}G^{\mu \nu} \rangle_{T,\mu} }{{\Lambda}^4}
      {}-{\pi}^3 \frac{896}{27} \frac{\alpha _s{\langle \bar{q}q
      \rangle }^2_{T,\mu} }{{\Lambda}^6}
   \right]\ .
\label{eq:WMC V}
\end{eqnarray}
Through these conditions
the dependences of the quark and gluonic condensates
on the temperature and/or the chemical potential
determine the intrinsic temperature and/or density
dependences 
of the bare parameters of the HLS Lagrangian,
which are then converted into 
those of the on-shell parameters through the Wilsonian RGEs.
As a result the parameters appearing in the hadronic 
thermal/dense-loop
corrections 
have the intrinsic temperature and/or density dependences:
$F_\pi$, $a$ and $g$ appearing there should be regarded as
\begin{eqnarray}
F_\pi &\equiv& F_\pi({\cal M}=0;T,\mu) \ ,
\nonumber\\
a &\equiv& a\left({\cal M}=M_\rho;T,\mu\right) \ ,
\nonumber\\
g &\equiv& g\left({\cal M}=M_\rho;T,\mu\right) \ ,
\end{eqnarray}
where the parametric mass
$M_\rho$ is determined from the on-shell condition:
\begin{equation}
   M_\rho^2 \equiv M_\rho^2(T,\mu) = 
   a({\cal M}=M_\rho ;T,\mu)
  g^2({\cal M}=M_\rho ;T,\mu)F_\pi^2({\cal M}=M_\rho ;T,\mu)\ .
\label{eq:Mdef}
\end{equation}

Now, let me consider the Wilsonian matching near the
chiral symmetry restoration point
with assuming that the quark condensate becomes zero
continuously for $(T,\mu) \rightarrow (T_c,\mu_c)$.
First, note that
the Wilsonian matching condition~(\ref{eq:WMC A}) 
provides
\begin{equation}
  \frac{F^2_\pi (\Lambda ;T_c,\mu_c)}{{\Lambda}^2} 
  = \frac{1}{8{\pi}^2}
  \left[
    1 + \frac{\alpha _s}{\pi}
    + \frac{2{\pi}^2}{3}
      \frac{\langle \frac{\alpha _s}{\pi} 
            G_{\mu \nu}G^{\mu \nu} \rangle_{T_c,\mu_c} }
      {{\Lambda}^4}
  \right]
 \neq 0 
\ ,
\label{eq:WMC A Tc}
\end{equation}
which implies that the matching with QCD dictates
\begin{equation}
F^2_\pi (\Lambda ;T_c,\mu_c) \neq 0 
\label{Fp2 Lam Tc}
\end{equation}
even at the critical point where the on-shell $\pi$
decay constant vanishes 
by adding the quantum corrections through
the RGE including the quadratic divergence~\cite{HY:conformal}
and hadronic thermal/dense-loop
corrections~\cite{HS:VMT,HKR}.
Second, we note that the axialvector and vector current correlators
$G_A^{(\rm{QCD})}$ and $G_V^{(\rm{QCD})}$
derived by the OPE
agree with each other for any value of $Q^2$.
Thus we require that
these current correlators in the HLS
in Eq.~(\ref{correlator HLS}) are
equal at the critical point
for any value of $Q^2\ \mbox{around}\ {\Lambda}^2$.
By taking account of the fact 
$F^2_\pi (\Lambda ;T_c,\mu_c) \neq 0$ derived from
the Wilsonian matching condition 
given in Eq.~(\ref{eq:WMC A Tc}),
the requirement 
$G_A^{(\rm{HLS})}=G_V^{(\rm{HLS})}$ is satisfied
only if the following conditions are met: 
\begin{eqnarray}
&&
g(\Lambda ;T,\mu) 
\ \mathop{\longrightarrow}_{(T,\mu) \to (T_c,\mu_c)}\ 
0\ , 
\label{eq:VMg}
\\
&&
a(\Lambda ;T,\mu) 
\ \mathop{\longrightarrow}_{(T,\mu) \to (T_c,\mu_c)}\ 
1 \ , 
\label{eq:VMa}
\\
&&
z_1(\Lambda ;T,\mu) - z_2(\Lambda ;T,\mu)
\ \mathop{\longrightarrow}_{(T,\mu) \to (T_c,\mu_c)}\ 
0\ . 
\label{eq:VMz}
\end{eqnarray}
As in Refs.~\citen{Brown-Rho:96,Brown-Rho:01b}, 
I expect that the quasiquark mass vanishes at the chiral restoration
point.
Since $m_q=0$ itself is a fixed point of the RGE for $m_q$ in 
Eq.~(\ref{rgewbaryons}),
$(g,a,m_q)=(0,1,0)$ is a fixed point of the coupled RGEs for $g$,
$a$ and $m_q$.
Then, the VM conditions $g=0$ and $a=1$ obtained for the bare
parameters
remain intact in the low-energy region including the on-shell
of $\rho$:
\begin{eqnarray}
&&
 g\,({\cal M} = M_\rho;T,\mu) 
\ \mathop{\longrightarrow}_{(T,\mu) \to (T_c,\mu_c)}\ 
0 \ ,
\nonumber\\
&&
 a\,({\cal M} = M_\rho;T,\mu) 
\ \mathop{\longrightarrow}_{(T,\mu) \to (T_c,\mu_c)}\ 
1 \ ,
\label{VM a g}
\end{eqnarray}
where the parametric $\rho$ mass 
$M_\rho=M_\rho(T,\mu)$ is determined from
the condition (\ref{eq:Mdef}).
The above conditions with Eq.~(\ref{eq:Mdef}) imply that
the parametric mass 
$M_\rho(T,\mu)$ also vanishes:
\begin{equation}
M_\rho(T,\mu) 
\ \mathop{\longrightarrow}_{(T,\mu) \to (T_c,\mu_c)}\ 
0 \ .
\label{VM Mrho}
\end{equation}

\section{Vector Manifestation in Hot and Dense Matter}
\label{sec:VMHDM}

In the previous section I have shown that the parametric $\rho$
mass $M_\rho$ vanishes at the chiral restoration point
due to the intrinsic temperature/density dependences obtained
from the Wilsonian
matching between the HLS and the OPE.
For obtaining the pole mass of $\rho$,
I need to included the hadronic thermal/dense-loop effects.
So far, 
the hadronic thermal/dense-loop corrections were calculated 
for two cases; $T>0$ with $\mu=0$~\cite{HS:VMT,HS:review}
and $\mu>0$ with $T=0$~\cite{HKR}.
In this section I summarize the resultant expressions of the
hadronic thermal and dense loop corrections to the $\rho$
pole mass in two cases, and then
show how the vector manifestation (VM)
is realized in hot
and dense matter.

\subsection{Vector meson mass in hot matter}
\label{ssec:VMMHM}

In Ref.~\citen{HS:VMT} 
the VM in hot
matter was shown to take place 
by using the hadronic thermal correction from the $\pi$
and $\rho$
to the $\rho$ pole mass
calculated in the
Landau gauge~\cite{HS:97}.
In Refs.~\citen{HKRS,HS:review}, the background field gauge was
adopted and the contribution from the quasiquark was included further.
Here, following Refs.~\citen{HKRS,HS:review}, I 
explain how the VM is realized in hot QCD.

I define pole masses of longitudinal and transverse modes of
$\rho$
from the poles of longitudinal and transverse components 
of the vector current correlator
at rest frame.
Hadronic thermal corrections from the $\pi$, $\rho$ and quasiquark
to the pole masses
were calculated at one-loop level in Ref.~\citen{HS:review},
and it was shown that, 
at the rest frame, the longitudinal pole mass agrees with the
transverse one.
The resultant expression for the $\rho$ pole mass is 
obtained as~\cite{HS:review}
\begin{eqnarray}
&&
  \left[m^L_\rho (T) \right]^2 =  \left[m^T_\rho (T) \right]^2 
  \equiv   m^2_\rho (T) 
\nonumber\\
&& \quad
= {M_\rho}^2
    +N_f\,g^2
    \Biggl[- \frac{a^2}{12}\tilde{G}_{2(B)}(M_\rho;T)
      + \frac{4}{5} \tilde{J}^2_{1(B)}(M_\rho;T)
      + \frac{33}{16} M_\rho^2 \tilde{F}^2_{3(B)}(M_\rho;M_\rho;T)
    \Biggr]
\nonumber\\
&& \qquad
  {}
   + N_c g^2 \, (1-\kappa)^2 
   \Biggl[ - \frac{4}{3}\tilde{J}_{1(F)}^2(m_q;T) -
     \frac{7m_q^2 - M_\rho^2}{6} \tilde{F}_{3(F)}^2(M_\rho;m_q;T)
   \Biggr]
\ ,
\label{eq:Mass} 
 \end{eqnarray}
where the explicit forms of the
functions expressing the bosonic corrections
[$\tilde{G}_{2(B)}$, $\tilde{J}^2_{1(B)}$ and 
$\tilde{F}^2_{3(B)}$] and fermionic corrections
[$\tilde{J}^2_{1(F)}$ and $\tilde{F}^2_{3(F)}$]
are listed in Appendix~\ref{app:Functions}.

Now, let me study the $\rho$ pole mass near the critical
temperature. 
As shown in section~\ref{sec:ITE}, the intrinsic temperature
dependences of the parameters of the HLS Lagrangian determined from
the Wilsonian matching imply that the
parametric $\rho$ mass $M_\rho$ vanishes at the critical
temperature. 
Furthermore, I expect that the quasiquark mass $m_q$ also vanishes
as was shown in, e.g., Refs.~\citen{Brown-Rho:96,Brown-Rho:01b}.
Then, near the critical temperature I should take 
$M_\rho \ll T$ and $m_q \ll T$ in Eq.~(\ref{eq:Mass}).
By noting that
\begin{eqnarray}
 \tilde{G}_{2(B)}(M_\rho ;T)
   &\stackrel{M_\rho \to 0}{\to}&\tilde{I}_{2(B)}(T)\ , \nonumber\\
 \tilde{J}^2_{1(B)}(M_\rho ;T)
   &\stackrel{M_\rho \to 0}{\to}&\tilde{I}_{2(B)}(T)\ , \nonumber\\
 {M_\rho}^2\tilde{F}^2_{3(B)}(M_\rho ;M_\rho ;T)
   &\stackrel{M_\rho \to 0}{\to}& 0, \nonumber\\
 \tilde{J}^2_{1(F)}(m_q;T)
   &\stackrel{m_q \to 0}{\to}& \tilde{I}_{2(F)}(T)\ , \nonumber\\
 m_q^2 \tilde{F}^2_{3(F)}(M_\rho;m_q;T)
   &\stackrel{M_\rho \to 0, m_q \to 0}{\to}& 0\ , \nonumber\\
 M_\rho^2 \tilde{F}^2_{3(F)}(M_\rho;m_q;T)
   &\stackrel{M_\rho \to 0, m_q \to 0}{\to}& 0\ ,
\end{eqnarray}
the pole mass of the vector meson
at $T \sim T_c$ becomes
\begin{eqnarray}
 m_\rho^2(T) 
  &=& {M_\rho}^2 +
      N_f\,g^2 \frac{15 - a^2}{144}T^2 +
      N_c\,g^2 (1 - \kappa)^2 \frac{1}{18}T^2. 
\label{eq:Mass2}
\end{eqnarray}
This shows that
the hadronic thermal effect gives a positive correction
in the vicinity of $a \simeq 1$,
and then the $\rho$ pole mass is actually larger than the
parametric mass $M_\rho$.
However, the intrinsic temperature dependences of the parameters
obtained in section~\ref{sec:ITE} lead to 
$g \to 0$ and $M_\rho \rightarrow0$ for $T \to T_c$.
Then, from  Eq.~(\ref{eq:Mass2}) it was
concluded~\cite{HS:VMT,HS:review} that
the $\rho$ pole mass $m_\rho$
vanishes at the critical temperature:
  \begin{equation}
   m_\rho (T) \to 0 \quad \mbox{for} \ T \rightarrow T_c \ .
  \end{equation}
This implies that the VM is realized 
at the critical temperature.

\subsection{Vector meson mass in dense matter}
\label{ssec:VMMDM}

In this subsection, following Ref.~\citen{HKR},
I briefly review how the VM
is realized in dense QCD.

In Ref.~\citen{HKR}
the pole masses of longitudinal and transverse modes of $\rho$
is defined from the poles of longitudinal and transverse
components of the vector current correlator
at rest frame, and it was shown that 
the hadronic dense loop corrections from the quasiquark to them 
agree with each other at one-loop level.
The resultant expression for the $\rho$ pole mass is expressed as
\begin{eqnarray}
&&
  \left[m^L_\rho (\mu) \right]^2 =  \left[m^T_\rho (\mu) \right]^2 
  \equiv   m^2_\rho (\mu) 
\nonumber\\
&& \quad
= M_\rho^2 
  + \frac{2}{3}\, g^2 (1-\kappa)^2
\Bigl[
  \bar{B}_S
  - (M_\rho^2 + 2 m_q^2) \,\mbox{Re}\,\bar{B}_0(M_\rho)
\Bigr]
\ ,
\label{pole mass dense}
\end{eqnarray}
where the functions
$\bar{B}_S$ and $\bar{B}_0(M_\rho)$ 
are defined in Eq.~(\ref{B fun}) in Appendix~\ref{app:Functions}.
Near $\mu \simeq \mu_c$ by taking $M_\rho\ll\mu$ and $m_q\ll\mu$,
this expression becomes
\begin{equation}
m_\rho^2(\mu) = M_\rho^2 + \frac{\mu^2}{6\pi^2} 
\left(1-\kappa\right)^2 \ ,
\label{pole mass dense 2}
\end{equation}
which
shows that the $\rho$ pole mass $m_\rho$
is larger than the parametric mass $M_\rho$ due to 
the hadronic dense-loop correction.
However, the intrinsic density dependences of the parameters
derived from the Wilsonian matching in section~\ref{sec:ITE}
imply that $g\rightarrow0$ and $M_\rho\rightarrow 0$ for
$\mu \rightarrow \mu_c$.
Then, from Eq.~(\ref{pole mass dense 2}), 
it was concluded~\cite{HKR} 
that the vector meson pole mass vanishes at the critical density:
\begin{equation}
m_\rho(\mu) \rightarrow 0 \quad
\mbox{for}\ \mu \rightarrow \mu_c \ .
\end{equation}
This implies that the VM is realized at the critical density.

\section{Summary}
\label{sec:summary}

In this talk I first 
summarized a main feature of the vector manifestation (VM)
proposed in Ref.~\citen{HY:VM} 
as a novel manifestation of Wigner realization of chiral 
symmetry in which the symmetry is restored at the critical point 
by the massless degenerate $\pi$ (pion and its flavor
partners) and $\rho$ ($\rho$ meson and its flavor partners) 
as the chiral
partner, 
in sharp contrast to the traditional manifestation {\' a} la 
the linear
sigma model where the symmetry is restored by the degenerate pion and
scalar meson.
Then, determining the intrinsic temperature and density dependences
of the bare parameters of the Lagrangian of the hidden local
symmetry (HLS) and including the hadronic thermal and dense
loop corrections to the $\rho$ pole mass,
I have shown how the VM is realized 
in hot and dense matter~\cite{HS:VMT,HKR,HKRS,HS:review}.
In this talk, I did not show the details of the calculations
on the hadronic thermal and dense loop corrections which 
can be seen in Refs.~\citen{HS:review} and \citen{HKR}

In Refs.~\citen{HS:VMT,HKRS,HS:review}, 
based on the VM in hot matter,
several predictions on the physical quantities were made:
the value of the critical temperature is expressed in terms
of the parameters in the OPE~\cite{HS:VMT};
the vector susceptibility actually agree with the axialvector 
susceptibility at the critical temperature taking non-zero 
value~\cite{HKRS};
the velocity of $\pi$ approaches the speed of light near the critical 
temperature~\cite{HKRS};
the vector dominance of the electromagnetic form factor
of pion is largely violated at the critical 
temperature~\cite{HS:review}; and so on.
I did not present those interesting results in this talk due to
the lack of time.

\section*{Acknowledgments}
I would like to thank Doctor Youngman Kim, 
Professor Mannque Rho, Doctor Chihiro Sasaki and Professor 
Koichi Yamawaki for collaboration in the works on which 
this talk is based.
I am very grateful to organizers for giving me an opportunity 
to present this talk.
This work was supported in part by the
Brain Pool program (\#012-1-44) provided by the Korean Federation
of Science and Technology Societies.

\appendix

\section{Functions}
\label{app:Functions}

In this Appendix, I list 
the integral forms of the
functions which appear in 
the expressions of hadronic thermal/dense-loop corrections to
the $\rho$ pole mass.

Let me list the functions used
in subsection~\ref{ssec:VMMHM} for expressing the
hadronic thermal-loop corrections.
The functions $\tilde{I}_{n(B,F)}(T)$ and
$\tilde{J}^n_{m(B,F)}(M ;T)$ ($n$, $m$: integers)
are given by
 \begin{eqnarray}
  &&\qquad\quad \tilde{I}_{n(B)}(T) 
   = \int \frac{d^3 k}{(2\pi)^3}\frac{|\vec{k}|^{n-3}}{e^{k/T}-1}
   = \frac{1}{2\pi^2}\hat{I}_{n(B)} T^n \ , \nonumber\\
  &&\qquad\quad \tilde{I}_{n(F)}(T) 
   = \int \frac{d^3 k}{(2\pi)^3}\frac{|\vec{k}|^{n-3}}{e^{k/T}+1}
   = \frac{1}{2\pi^2}\hat{I}_{n(F)} T^n\ , \nonumber\\
  &&\qquad\qquad\qquad \hat{I}_{2(B)} = \frac{{\pi}^2}{6},\quad
  \hat{I}_{4(B)} = \frac{{\pi}^4}{15}\ , \quad
  \hat{I}_{2(F)} = \frac{\pi^2}{12}\ , 
\label{I fun}
\\
  &&\quad \tilde{J}^n_{m(B)}(M ;T) 
   = \int \frac{d^3 k}{(2\pi)^3} \frac{1}{e^{\omega /T}-1}
      \frac{|\vec{k}|^{n-2}}{{\omega}^m}\ ,
\nonumber\\
  &&\quad \tilde{J}^n_{m(F)}(M;T) 
   = \int \frac{d^3 k}{(2\pi)^3} \frac{1}{e^{\omega /T}+1}
      \frac{|\vec{k}|^{n-2}}{{\omega}^m}\ , 
\label{J fun}
\end{eqnarray}
where
The functions $\tilde{F}^n_{3(B,F)}(p_0;M;T)$ and
$\tilde{G}_{n(B)}(p_0;T)$ are defined as
\begin{eqnarray}
  &&\tilde{F}^n_{3(B)}(p_0;M;T) 
   = \int \frac{d^3 k}{(2\pi)^3}\frac{1}{e^{\omega /T}-1}
      \frac{4|\vec{k}|^{n-2}}{\omega (4{\omega}^2 - {p_0}^2)}\ , 
\nonumber\\
  &&\tilde{F}^n_{3(F)}(p_0;M;T) 
   = \int \frac{d^3 k}{(2\pi)^3}\frac{1}{e^{\omega /T}+1}
      \frac{4|\vec{k}|^{n-2}}{\omega (4{\omega}^2 - {p_0}^2)}\ ,
\label{D.3}
\\
  &&\quad \tilde{G}_{n(B)}(p_0;T) 
   = \int \frac{d^3 k}{(2\pi)^3}\frac{|\vec{k}|^{n-3}}{e^{k/T}-1}
       \frac{4|\vec{k}|^2}{4|\vec{k}|^2 - {p_0}^2}\ .
\label{F G fun}
\end{eqnarray}

The functions appearing in the hadronic dense loop correction
from the quasiquark to the $\rho$ pole mass are given by
\begin{eqnarray}
\bar{B}_S &=& 
\frac{1}{4\pi^2} \left[P_F \omega_F - m_q^2 \,
\ln \frac{ P_F + \omega_F }{m_q} \right] \ ,
\nonumber\\
  \bar{B}_0(M)
&=&
  \frac{1}{8\pi^2}
  \Biggl[
    - \ln \frac{ P_F + \omega_F }{ m_q }
    {}+ \frac{1}{2}
    \sqrt{ 
      \frac{ 4m_q^2 -M^2 - i \epsilon }{ -p_0^2 - i \epsilon } 
    }
\nonumber\\
&& \qquad\times
    \ln \frac{
      \omega_F\,\sqrt{ 4m_q^2 -M^2 - i \epsilon }
      + P_F\,\sqrt{ -M^2 - i \epsilon }
    }{
      \omega_F\,\sqrt{ 4m_q^2 -M^2 - i \epsilon }
      - P_F\,\sqrt{ -M^2 - i \epsilon }
    }
  \Biggr]
\ ,
\label{B fun}
\end{eqnarray}
where 
$P_F$ is the Fermi momentum of the quasiquark and 
$\omega_F \equiv \sqrt{ P_F^2 + m_q^2}$.
Note that, in the present analysis, I can take 
$P_F = \sqrt{\mu^2 - m_q^2}$ and $\omega_F = \mu$.

\end{document}